\documentclass{article}

\usepackage[dblblindworkshop, final, nonatbib]{neurips_2025_libribrain_competition}

\workshoptitle{2025 PNPL Competition}

\usepackage[utf8]{inputenc} 
\usepackage[T1]{fontenc}    
\usepackage{hyperref}       
\usepackage{url}            
\usepackage{booktabs}       
\usepackage{amsfonts}       
\usepackage{nicefrac}       
\usepackage{microtype}      
\usepackage{xcolor}         
\usepackage{graphicx}
\usepackage{amsmath}

\title{MEBM-Phoneme: Multi-scale Enhanced BrainMagic for End-to-End MEG Phoneme Classification}

\author{%
  Liang Jinghua$^{1}$ \And Zhang Zifeng$^{1,2}$ \And Li Songyi$^{1}$ \And Zheng Linze$^{1}$
  \\
  $^{1}$Speech and Hearing Research Center, School of Intelligence Science and Technology \\
  $^{2}$Center for BioMed-X Research, Academy for Advanced Interdisciplinary Studies \\ 
  Peking University, Beijing, China 100871\\
  \texttt{zifengzhang25@stu.pku.edu.cn} \\
}

\begin{document}

\maketitle

\begin{abstract}
We propose \textbf{MEBM-Phoneme}, a multi-scale enhanced neural decoder for phoneme classification from non-invasive magnetoencephalography (MEG) signals.
Built upon the BrainMagic backbone, MEBM-Phoneme integrates a short-term multi-scale convolutional module to augment the native mid-term encoder, with fused representations via depthwise separable convolution for efficient cross-scale integration.
A convolutional attention layer dynamically weights temporal dependencies to refine feature aggregation.
To address class imbalance and session-specific distributional shifts, we introduce a stacking-based local validation set alongside weighted cross-entropy loss and random temporal augmentation.
Comprehensive evaluations on \textbf{LibriBrain Competition 2025 Track~2} demonstrate robust generalization, achieving competitive phoneme decoding accuracy on the validation and official test leaderboard.
These results underscore the value of hierarchical temporal modeling and training stabilization for advancing MEG-based speech perception analysis.
\end{abstract}

\section{Introduction}
\label{introduction}
Phoneme decoding from brain signals has long been a central goal of neural speech decoding research. 
Recent advances in invasive neuroprosthetic technologies achieve remarkable accuracy by directly mapping neural activity to phoneme categories and then decoding text through language models~\cite{metzger2023,willett2023,card2024}.
However, replicating such performance using non-invasive neuroimaging techniques, such as MEG, remains highly challenging due to the lower signal-to-noise ratio.

To address this problem, we propose MEBM-Phoneme, an enhanced end-to-end framework for MEG-based phoneme classification. 
Our method is developed for Track 2 of the NeurIPS 2025 LibriBrain Competition \cite{landau2025,ozdogan2025}, which focuses on decoding phonemic representations from non-invasive MEG recordings. 

Our approach centers on three key contributions:
\begin{enumerate}
    \item \textbf{Model Architecture:} We augment the BrainMagic~\cite{defossez2023} architecture with a short-term multi-scale convolutional module, capturing fine-grained temporal dependencies. 
    The resulting features are fused with mid-term representations through a depthwise separable convolution, followed by a convolutional attention layer that aggregates temporal information.
    
    \item \textbf{Validation Strategy:} To address severe class imbalance and better approximate the holdout distribution, we construct a session-aware local validation set using a stacking-based sampling method, ensuring statistical alignment with the competition’s evaluation protocol.
    
    \item \textbf{Training Protocol:} To enhance robustness and address class imbalance, we adopt a stochastic sample construction strategy that randomly selects a phoneme class per iteration and dynamically averages a variable number of instances. Together with random temporal offsets and an adaptive weighted cross-entropy loss, this approach promotes balanced learning and stable convergence.
\end{enumerate}

\section{Methods}
\label{methods}

Our approach for MEG-based phoneme classification is designed to enhance temporal feature modeling, handle severe class imbalance, and improve training robustness. Below, we present the detailed description of our method.

\subsection{Model Architecture}
\begin{figure}[t]
  \centering
  \includegraphics[width=0.9\linewidth]{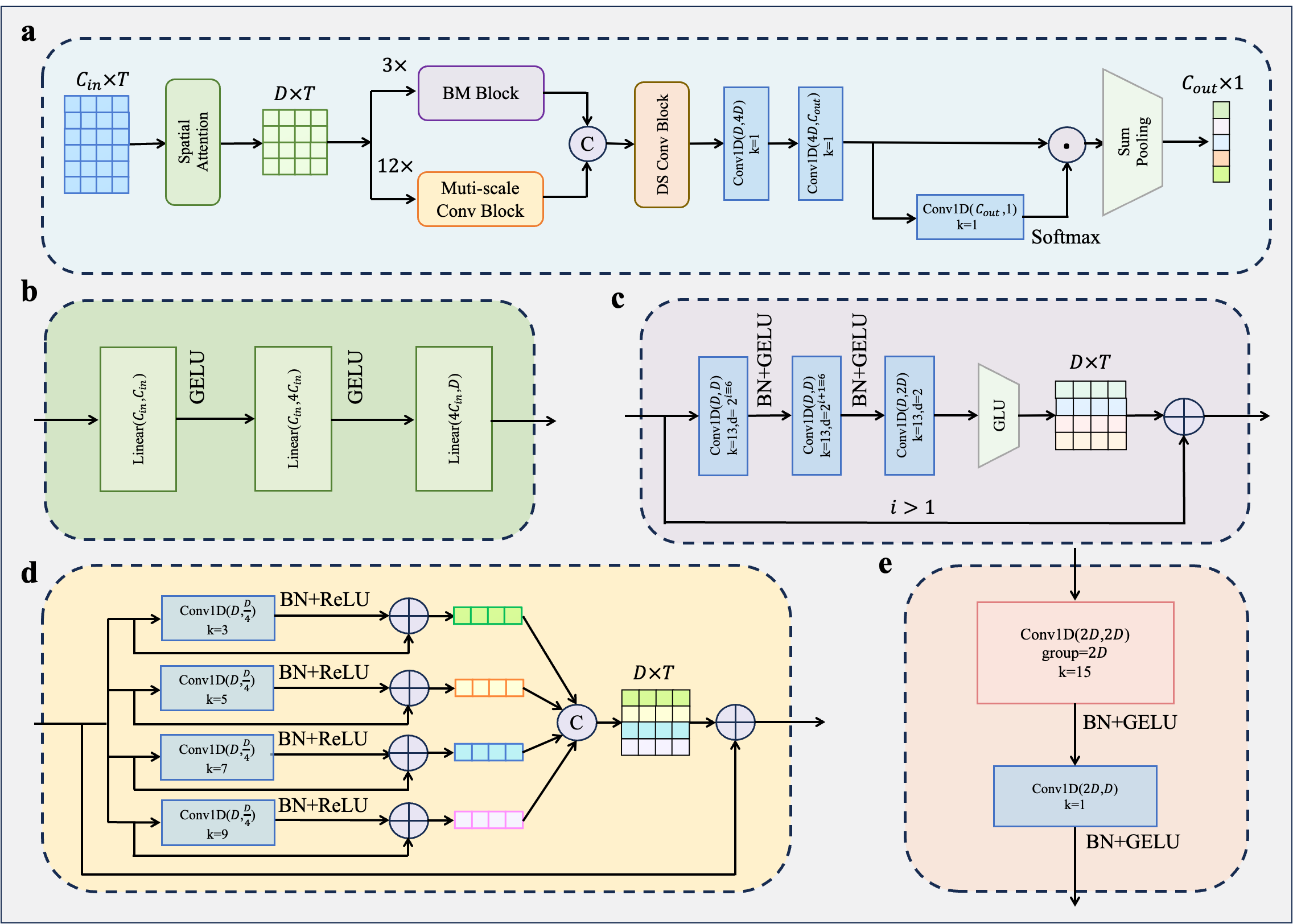}
  \caption{
  Overall architecture of the proposed MEBM-Phoneme model.  
  \textbf{(a)} The complete processing pipeline.
  \textbf{(b)} The \textit{spatial attention module} enhances sensor-level representations by learning spatial relevance weights across MEG channels.  
  \textbf{(c)} The \textit{BM encoder} extracts mid-term contextual features from spatially weighted signals.  
  \textbf{(d)} The \textit{short-term multi-scale convolutional module} captures fine-grained temporal dependencies using multiple receptive fields.  
  \textbf{(e)} The \textit{depthwise separable convolutional layer} further refines temporal representations with lightweight channel-wise and pointwise filtering.
  }
  \label{fig:model}
\end{figure}
As illustrated in Figure~\ref{fig:model}, our proposed MEBM-Phoneme model builds upon the original BrainMagic architecture by introducing a dedicated short-term feature extraction pathway and an enhanced fusion mechanism.
Given the MEG input $\mathbf{X} \in \mathbb{R}^{C_{\text{in}} \times T}$, where $C_{\text{in}}$ denotes the number of MEG sensor channels and $T$ represents the total number of temporal samples, the model first applies a spatial attention module that dynamically re-weights sensor-wise activations, producing a spatially enhanced representation $\mathbf{H}_s \in \mathbb{R}^{D \times T}$, with $D$ being the dimensionality of the projected feature space.
This representation is then processed in parallel by two temporal streams:
12 multi-scale convolutional blocks comprising a stack of dilated convolutional blocks designed to capture local temporal dependencies across multiple receptive fields, and
3 BrainMagic (BM) encoders responsible for extracting mid-term contextual features.
The outputs from both branches are concatenated along the channel dimension and passed through a depthwise separable convolution for efficient fusion.
This operation not only reduces computational overhead but also enforces feature disentanglement across temporal scales.
Subsequently, a convolutional attention layer aggregates temporal information through a channel-compressive operation. 
Specifically, a $1$D convolution reduces the feature dimension to $1$, yielding an attention map $\mathbf{A} \in \mathbb{R}^{1 \times T}$. 
A softmax normalization is then applied along the temporal axis to obtain attention weights $\mathbf{W}_t$, which are multiplied back with the fused representation to reweight each time step by its learned importance:
\[
\mathbf{H}_{\text{att}} = \mathbf{W}_t \odot \mathbf{H}_{\text{fused}}.
\]
Finally, a sum pooling operation collapses the temporal dimension, and a linear layer with softmax activation produces the phoneme-level class probabilities.  

\subsection{Validation and Training Sampling Strategy}
\label{sec:val_train_strategy}
To ensure robust and distributionally aligned evaluation, we design a unified data construction rule for both validation and training samples, differing only in the degree of stochasticity.  
For each phoneme class, we estimate the average number of samples per session $n$, and determine the number of averaged single samples $n'$ according to:

\begin{equation}
n'_{\text{val}} =
\begin{cases}
100, & n > 100,\\
n, & 50 \le n < 100,\\
1.5n, & n < 50,
\end{cases}
\label{eq:val_rule}
\end{equation}

\begin{equation}
n'_{\text{train}} =
\begin{cases}
100, & n > 100,\\
\mathrm{rand}[n-5, \min(n+5, 100)], & 50 \le n < 100,\\
2n, & n < 50.
\end{cases}
\label{eq:train_rule}
\end{equation}

Here, $n'_{\text{val}}$ defines a deterministic sampling rule for validation, while $n'_{\text{train}}$ introduces controlled randomness during training to enhance generalization and reduce overfitting.  
At each training iteration, a single phoneme class is randomly selected, and its samples are averaged following Eq.~\ref{eq:train_rule}.  
To further improve temporal robustness, we apply a random temporal jittering scheme: the starting point of each segment is uniformly sampled from the interval $[\text{onset}-3, \text{onset}+3]$, and a fixed $0.5$\,s window is subsequently extracted.  
This perturbation increases invariance to onset timing variability inherent in MEG signals.  

Finally, training is guided by an adaptive weighted cross-entropy loss, designed not only for class balancing but also to reduce confusion among acoustically or articulatorily similar phonemes.  

\section{Experiments}
\label{sec:experiments}

\subsection{Experimental Setup}
\label{sec:setup}
The offline validation set was constructed using the official validation and test sessions (\textit{Sherlock1, sessions 11–12}) to approximate the holdout distribution defined by the LibriBrain challenge.  
For reproducibility, we fixed random seeds and performed eight independent sampling iterations per phoneme class, discarding classes with insufficient samples to meet the required $n'$ values from Eq.~\ref{eq:val_rule}.  
This procedure ensured that the resulting validation data statistically aligned with the holdout distribution while mitigating class imbalance and session-specific bias.

Before training, the continuous MEG signals of each session were normalized along the temporal dimension independently.  
After sample extraction and averaging, the resulting averaged samples were normalized again along the temporal axis.  

The proposed MEBM-Phoneme model was implemented in PyTorch and trained on a single NVIDIA A800 GPU (80\,GB) for approximately three hours.  
The network contained 4.7\,M trainable parameters.  
Each input MEG sequence consisted of $C_{\text{in}} = 306$ channels and $T = 125$ time points, producing $C_{\text{out}} = 39$ phoneme probabilities.  
The intermediate feature dimension was set to $D = 128$ with a dropout rate of 0.02.  
Training was conducted for 80 epochs using the AdamW optimizer with a learning rate of $1\times10^{-3}$, batch size of 256, and 40,000 samples per epoch.  
All convolutional layers adopted \texttt{padding='same'} to preserve temporal resolution.  
Model selection and hyperparameter tuning were performed using the offline validation set constructed from the \textit{Sherlock1 Session 11–12} data.

\subsection{Results and Ablation}
\label{sec:results}

We report the performance of the proposed MEBM-Phoneme model and its ablated variants on the offline validation set.  
All results are averaged over six random seeds $\{0,1,2,3,4,5\}$ for reproducibility.  
Evaluation metrics include F1\textsubscript{macro}(\%), Top-3 Acc\textsubscript{macro} (\%), and Top-5 Acc\textsubscript{macro} (\%).
Table~\ref{tab:ablation} summarizes the performance of our proposed MEBM-Phoneme model and its ablated variants on the validation set. 
The full model achieves an average F1\textsubscript{macro} of 60.95\%, Top-3 Acc\textsubscript{macro} of 89.54\%, and Top-5 Acc\textsubscript{macro} of 95.08\% across six random seeds. Removing any individual component leads to a consistent degradation in performance, demonstrating the effectiveness of each module in the proposed architecture.
Moreover, all model variants maintain relatively high Top-3 and Top-5 Acc\textsubscript{macro} scores, indicating that even when the top prediction is incorrect, the correct phoneme often lies among the top few candidates.  
This suggests that the model already possesses a strong discriminative capacity for phoneme categorization, and could further benefit from integration with a language model to leverage contextual linguistic information.

On the online test set, our model achieved a peak decoding accuracy of up to 72\% on the first half, but exhibited degraded performance on the second half. We conjecture that this discrepancy may be partly attributed to our submission strategy. Nevertheless, results on the local evaluation set indicate that our approach remains robust and demonstrates strong generalization capability.

\begin{table}[t]
\centering
\caption{Results and ablation analysis on the local validation set under six random seeds (0–5). 
Metrics include F1\textsubscript{macro}, Top-3 Acc\textsubscript{macro}, and Top-5 Acc\textsubscript{macro} (mean ± std).}
\label{tab:ablation}
\begin{tabular}{lccc}
\toprule
\textbf{Model Variant} & \textbf{F1\textsubscript{macro} (\%)} & \textbf{Top-3 Acc\textsubscript{macro} (\%)} & \textbf{Top-5 Acc\textsubscript{macro} (\%)} \\
\midrule
Full Model     & \textbf{60.95±0.90} & \textbf{89.54±0.48} & \textbf{95.08±0.61} \\
w/o Weighted Loss               & 59.97±0.90 & 88.87±1.14 & 94.75±0.63 \\
w/o Multi-scale Conv            & 59.75±0.68 & 88.98±1.12 & 94.67±1.03 \\
w/o BM Encoder                  & 54.43±2.07 & 84.96±1.69 & 92.19±1.28 \\
w/o Conv. Attention             & 59.60±0.82 & 88.47±1.46 & 94.17±1.13 \\
\bottomrule
\end{tabular}
\end{table}

\section{Conclusion}
\label{sec:conclusion}

This work presents MEBM-Phoneme, our enhanced framework for MEG-based phoneme classification in the NeurIPS 2025 LibriBrain Competition. 
By augmenting the BrainMagic architecture with a short-term multi-scale convolutional module and an attention-based temporal aggregation mechanism, 
the model effectively captures both fine-grained and contextual temporal dependencies from non-invasive MEG signals. 
Additionally, our session-aware validation strategy and stochastic training protocol improve robustness against class imbalance and distributional variation. 

Experimental results under multiple random seeds demonstrate that each component of MEBM-Phoneme contributes to stable performance improvements, achieving competitive results on the official evaluation set.
It is important to note, however, that our study relies on averaged MEG signals to boost the signal-to-noise ratio. A significant remaining challenge—and the focus of our future work—is to perform accurate phoneme classification on single-trial, continuous MEG data, which is essential for developing practical, real-time neural speech decoding systems.

\begin{ack}
This work was supported by the STI 2030—Major Projects (No. 2021ZD0201500), the High-performance Computing Platform of Peking University, and the Biomedical Computing Platform of National Biomedical Imaging Center of Peking University. We also gratefully acknowledge the guidance and valuable discussions provided by Prof. Jing Chen.
\end{ack}


\appendix

\section*{Appendix A. Adaptive Loss Weights for Phoneme Classes}

Table~\ref{tab:loss_weights} lists the adaptive loss weights used for each phoneme class. 
The weights were empirically tuned to balance class frequency and confusion. The remaining phoneme weights were all set to 1.0.

\begin{table}[h]
\centering
\caption{Adaptive loss weights for each phoneme class.}
\label{tab:loss_weights}
\begin{tabular}{lcccccccccc}
\toprule
\textbf{Phoneme} & /ey/ & /ay/ & /uh/ & /uw/ & /s/ & /sh/ & /m/ & /ae/ & /jh/ & /ah/ \\ 
\midrule
\textbf{Weight}  & 0.05 & 3.00 & 10.00 & 3.00 & 0.80 & 3.00 & 3.00 & 3.00 & 1.50 & 2.00 \\
\bottomrule
\end{tabular}
\end{table}



\medskip
{
\small
\begin{thebibliography}{9}

\bibitem{metzger2023}
Metzger, S.L., Littlejohn, K.T., Silva, A.B., Moses, D.A., Seaton, M.P., Wang, R., Dougherty, M.E., Liu, J.R., Wu, P., Berger, M.A., Zhuravleva, I., Tu-Chan, A., Ganguly, K., Anumanchipalli, G.K.\ \& Chang, E.F.\ (2023)
A high-performance neuroprosthesis for speech decoding and avatar control. {\it Nature}, {\bf 620}(7976), 1037--1046.

\bibitem{willett2023}
Willett, F.R., Kunz, E.M., Fan, C., Avansino, D.T., Wilson, G.H., Choi, E.Y., Kamdar, F., Glasser, M.F., Hochberg, L.R., Druckmann, S., Shenoy, K.V.\ \& Henderson, J.M.\ (2023) 
A high-performance speech neuroprosthesis. {\it Nature}, {\bf 620}(7976), 1031--1036.

\bibitem{card2024}
Card, N.S., Wairagkar, M., Iacobacci, C., Hou, X., Singer-Clark, T., Willett, F.R., Kunz, E.M., Fan, C., Vahdati Nia, M., Deo, D.R., Srinivasan, A., Choi, E.Y., Glasser, M.F., Hochberg, L.R., Henderson, J.M., Shahlaie, K., Stavisky, S.D.\ \& Brandman, D.M.\ (2024)
An accurate and rapidly calibrating speech neuroprosthesis.
{\it New England Journal of Medicine}, {\bf 391}(7):609--618.

\bibitem{landau2025}
Landau, G., Özdogan, M., Elvers, G., Mantegna, F., Somaiya, P., Jayalath, D., Kurth, L., Kwon, T., Shillingford, B., Farquhar, G., Jiang, M., Jerbi, K., Abdelhedi, H., Ramos, Y.M., Gulcehre, C., Woolrich, M., Voets, N.\ \& Jones, O.P.\ (2025)
The 2025 PNPL Competition: Speech Detection and Phoneme Classification in the LibriBrain Dataset.
{\it arXiv}, 2506.10165.
doi:10.48550/arXiv.2506.10165.

\bibitem{ozdogan2025}
Özdogan, M., Landau, G., Elvers, G., Jayalath, D., Somaiya, P., Mantegna, F., Woolrich, M.\ \& Jones, O.P.\ (2025) LibriBrain: Over 50 hours of within-subject MEG to improve speech decoding methods at scale.
{\it arXiv}, 2506.02098.
doi:10.48550/arXiv.2506.02098.

\bibitem{defossez2023}
Défossez, A., Caucheteux, C., Rapin, J., Kabeli, O.\ \& King, J.-R.\ (2023)
Decoding speech perception from non-invasive brain recordings.
{\it Nature Machine Intelligence}, {\bf 5}(10):1097--1107.


\end{thebibliography}
}
\end{document}